\documentclass{article}
\usepackage[utf8]{inputenc}
\DeclareUnicodeCharacter{202F}{\,} % map narrow no-break space to thin space
\usepackage[T1]{fontenc}
\usepackage{hyperref}
\usepackage{xurl}
\usepackage[colorinlistoftodos, color=lightgray]{todonotes}
\usepackage{graphicx} % Required for inserting images
\usepackage{pifont} % for tick
\usepackage{threeparttable}
\usepackage{cancel}
\usepackage{array}
\usepackage{colortbl}
\usepackage{tikz}
\usepackage{adjustbox}
\usepackage{tabularx}
\usepackage{booktabs}
\usepackage[preprint]{tmlr}

\title{Introducing the Generative Application Firewall (GAF)}

\author{%
\vspace{-20pt}%
\begin{center}
\begin{minipage}{\textwidth}
\centering
\vspace{6mm}
\begin{tabular}{l@{\hspace{2cm}}l}
  \name Joan Vendrell Farreny\addr NeuralTrust & \\
  \name Martí Jordà Roca \addr NeuralTrust & \\
  \name Miquel Cornudella Gaya \addr NeuralTrust & \\
  \name Rodrigo Fernández Baón \addr NeuralTrust & \\
  \name Victor García Martínez\addr NeuralTrust & \\
  \name Eduard Camacho Sucarrats \addr NeuralTrust & \\
  \name Alessandro Pignati \addr NeuralTrust & \\[6mm]
  \textbf{Expert endorsers:} & \\[0.5em]
  \name Christos Kalyvas-Kasopatidis$^*$  \addr University of the Aegean & \\
  \name Emre Isik$^*$  \addr University of Cambridge & \\
  \name Idan Habler$^*$  \addr OWASP GenAI Security Project & \\
  \name Jinwei Hu$^*$ \addr University of Liverpool & \\
  \name Strahinja Janjusevic$^*$  \addr MIT Computer Science and Artificial Intelligence Laboratory  & \\
  \name Osvaldo Ramirez$^*$ \addr Center for AI and Digital Policy & \\
  \name Tu Nguyen$^*$  \addr Huawei & \\
  \name Xuanlin Liu$^*$  \addr Cloud Security Alliance  & \\
\end{tabular}
\end{minipage}
\end{center}
}

\begin{document}
 
\maketitle
\def\thefootnote{*}\footnotetext{The paper represents the views of the individual endorsers and not necessarily of their affiliated institutions.}
\def\thefootnote{\arabic{footnote}}
\vspace{-10pt}
\begin{abstract}
This paper introduces the Generative Application Firewall (GAF), a new architectural layer for securing LLM applications. Existing defenses---prompt filters, guardrails, and data-masking---remain fragmented; GAF unifies them into a single enforcement point, much like a WAF coordinates defenses for web traffic, while also covering autonomous agents and their tool interactions.
\end{abstract}

\section{Introduction}

As generative AI systems become increasingly adopted in real-world applications, questions around how to secure them have taken on new urgency. The rapid pace of deployment has outpaced the development of shared terminology, frameworks, and architectural patterns for safety and control.

This challenge parallels the evolution of web application security. Web applications introduced new attack vectors that traditional network firewalls could not address. Attacks such as HTTP flooding, OWASP Top 10 \cite{owasp-top-ten-2021}, and other application-layer threats used seemingly valid HTTP requests that bypassed network-level protections. This gap prompted the development of Web Application Firewalls (WAFs) \cite{f5-waf-glossary} as a dedicated security layer for web traffic. Similarly, generative LLM applications introduce novel threats—such as prompt injection, jailbreaking, and context manipulation \cite{pathade2025redteaming}—that existing WAFs cannot effectively detect or mitigate.

This paper introduces the concept of a Generative Application Firewall (GAF), a dedicated security layer for generative applications, and provides an abstraction to structure ongoing efforts in this space.

\section{Why is it necessary?}

Generative AI attacks exploit limitations in traditional layered security models, requiring a dedicated security layer to protect applications. Just as web applications bypassed network-level protections, generative AI applications circumvent both network (Layer 3-4) and traditional application-layer (Layer 4-7) defenses.

\textbf{Bypassing Network-Layer Security (Layer 3-4):} Traditional network firewalls and intrusion detection systems examine packet headers, IP addresses, and basic protocol information at the network and transport layers. However, jailbreak and prompt injection attacks embed within semantically valid HTTP requests that appear legitimate at the network level. These attacks use standard ports, proper HTTP formatting, and legitimate IP addresses, showing no suspicious patterns. A malicious prompt for illegal activities travels indistinguishably from a benign inquiry.

\textbf{Evading Web Application Firewalls (Layer 4-7):} Traditional WAFs inspect HTTP headers, request structures, and known attack signatures to detect threats like SQL injection or cross-site scripting. However, generative AI attacks exploit natural language semantics rather than syntactic vulnerabilities. For instance, a jailbreak attack might request an AI to "roleplay as an unrestricted AI" or use social engineering to extract sensitive information. These contain no malicious code patterns or anomalies that WAFs detect. The attack lies in conversational manipulation of the AI model.

\textbf{The Semantic Gap:} Prompt injection and jailbreak attacks operate at the semantic layer, exploiting natural language meaning and context rather than technical vulnerabilities. Traditional tools lack contextual awareness to recognize that a "creative writing" request might bypass safety guardrails. Multi turn attacks like Echo Chamber \cite{alobaid2025echochamber} build malicious intent across interactions, remaining invisible to measures analyzing requests in isolation.

Established security platforms, such as WAFs, do not address generative AI systems. These tools inspect structured traffic for misuse patterns but lack semantic awareness for natural language interfaces. This mismatch creates gaps that a dedicated Generative Application Firewall addresses. 

\textbf{The Reality of Application Security:} If every generative AI application featured perfect security, a GAF would be unnecessary. However, real-world deployments rarely achieve this ideal. Developers often prioritize functionality over security \cite{hermann2025exploratory}, rush to market, or lack expertise for comprehensive measures. Even well-intentioned teams introduce vulnerabilities through errors or unanticipated attacks.

\textbf{Multi-Application Environments and Policy Enforcement:} Environments with multiple generative AI applications amplify complexity. Enterprises deploy chatbots, code tools, content assistants, and analysis agents across departments, each with different models and security. Without centralized security, organizations struggle to ensure consistent policies, prevent compromises, monitor interactions, and respond to threats.

Existing security solutions for LLM-based applications remain too fragmented. Prompt filters stop some injection attempts, data-masking layers redact sensitive inputs and outputs, and guardrails confine an application's functional scope. Each technique is valuable and necessary, yet together they fall short of providing the level of protection that is standard in traditional web applications. What is still missing is a unifying architectural layer that, much as a Web Application Firewall (WAF) coordinates multiple defenses for conventional web traffic, can orchestrate these disparate controls, fill the remaining gaps, and offer engineers a single point of policy enforcement. We refer to this missing layer as the Generative Application Firewall.

\section{Generative Application Firewall}

A Generative Application Firewall (GAF) is a security and control layer designed for applications that expose a natural language interface powered by large language models. It sits between the user and the underlying LLM application, serving as a centralized enforcement point for security policies across all layers of the generative application stack: from network-level controls like rate limiting and access restrictions to semantic-level filtering and context-aware policy enforcement.

Unlike prompt-level guardrails, the GAF maintains a holistic view of the application, spanning users, sessions, and interactions over time. Its goal is to unify fragmented protections into a coherent, enforceable perimeter for generative systems.

Beyond user-to-LLM prompts, GAF mediates autonomous agents by intercepting agent-initiated tool calls, enforcing per-tool policies, and filtering tool outputs to prevent indirect injection and data poisoning.

\paragraph{Motivating example.} A user attempts an indirect jailbreak by asking the assistant to "roleplay as a compliance auditor" and then requests detailed steps for constructing an improvised explosive. The Semantic layer flags intent and either blocks or selectively redacts; if the user escalates across turns with euphemisms, the Context layer detects the pattern and terminates streaming mid-generation while logging the event for audit.

\subsection{Threat Model}

GAF protects conversational systems by enforcing policies across inputs, outputs, tools, and longitudinal context. We consider the following elements:

\textbf{Assets.} Application tools and actions; retrieval and knowledge bases; user data and PII within context windows; credentials (API keys, OAuth tokens); security policies and allow/deny lists; logs and telemetry; user identities, roles, and rate limits; agent plans/state, tool call graphs, and tool outputs (treated as untrusted inputs).

\textbf{Adversaries.} (1) External, unauthenticated actors attempting prompt injection, jailbreaks, scraping, model extraction, and DoS via volumetric or semantic means; (2) Authenticated but malicious or compromised users (insider threats) escalating privileges or abusing sensitive tools; (3) Automated probing and coordinated bots exploring multi‑turn bypass strategies; (4) Content supply‑chain adversaries poisoning retrieved context or prompts through indirect injection; autonomous agents executing unsafe or unbounded tool sequences; poisoned tool outputs used for indirect injection.

\textbf{Trust boundaries.} GAF operates at multiple trust boundaries where security policy must be enforced: End user \(\leftrightarrow\) GAF; GAF \(\leftrightarrow\) application backend; GAF \(\leftrightarrow\) LLM/provider; GAF \(\leftrightarrow\) data stores/RAG; GAF \(\leftrightarrow\) tool/executor sandboxes; streaming channels (WebSocket/SSE/gRPC) where mid‑generation control is required.

\textbf{Assumptions.} Channels are protected by TLS; identity providers are authoritative for authentication; GAF has inline authority to allow, block, redact, redirect, alert, or terminate streams; underlying applications may contain latent vulnerabilities; all decisions and events are auditable.

\textbf{Security goals.} Confidentiality (prevent sensitive data exfiltration); integrity (prevent policy‑breaking tool actions, prompt/response tampering, and indirect injection or data poisoning via tool outputs); availability (maintain latency/throughput under attack); compliance and auditability (traceable decisions). Attacks span single‑turn semantic prompts and multi‑turn context‑building strategies; GAF addresses both with the Semantic and Context layers.

\subsection{Security Layers} \label{sec:layers}

A core principle of the Generative Application Firewall (GAF) is that securing natural-language interfaces requires protection across several interdependent layers. This layered approach is an application of the defense-in-depth principle, a well-established security concept now being adapted for LLM agents~\cite{zhang2025llm}. Each layer addresses a distinct class of risks, and only in combination do they provide comprehensive coverage.

This framework defines security layers for GAF, each targeting specific threats and extending traditional models into the generative AI domain:

\begin{enumerate}

\item \textbf{Network Layer}. Network-level protection enforces request and prompt rate limits, blocks abusive IPs, and restricts traffic. These controls mitigate prompt-flooding, model extraction, denial-of-service, scraping, and other volumetric abuse before users are identified or sessions are established. By shaping inbound volume and excluding known-bad sources at the perimeter, this layer reduces background noise and protects upstream components from saturation.

\item \textbf{Access Layer}. Access-level protection ensures users are authenticated, authorized, and assigned appropriate roles. It defends against session hijacking, privilege escalation, and misuse of sensitive capabilities by authenticated but malicious actors. Practically, this layer maps organizational policies (e.g., least-privilege for tools, retrieval scopes, or model actions) onto identities and sessions so that sensitive operations are gated by principled access decisions, including per-tool scopes and allow/deny lists for agents.

\item \textbf{Syntactic Layer}. The syntactic layer inspects the structure of inputs and outputs to catch escape sequences, obfuscated tokens, and code injections that could impact downstream components such as databases, interpreters, or execution environments. It validates formats, tool invocation schemas, and return payload formats, enforces type/length/encoding constraints, and monitors message framing in both discrete requests and streamed responses. When payloads are dangerous due to how they're encoded rather than what they mean, the syntactic layer is the choke point.

\item \textbf{Semantic Layer}. This layer is responsible for detecting and preventing semantic attacks that exploit the meaning and context of natural language interactions. It includes mechanisms for detecting prompt encodings or single-turn attacks like DAN \cite{dan-medium-2022}, DEV mode \cite{kirch2025jailbreak}, or Best-of-N jailbreak \cite{hughes2024bestofnjailbreak} or other semantic-level threats. These are attacks where detection doesn't require understanding the broader context of the model's use case or conversation history; the malicious intent can be identified from the individual prompt alone, including injection in agent tool parameters and prompts.

\item \textbf{Context Layer}. Context-level protection addresses adversarial multi-turn jailbreaks that evolve across a dialogue. By maintaining session history and role awareness, it detects strategies such as Echo Chamber \cite{alobaid2025echochamber} and Crescendo \cite{russinovich2024crescendo}, where malicious instructions are incrementally introduced to bypass guardrails. This requires correlating prompts and outputs over time, surfacing semantic buildup, and enforcing longitudinal policies such as exfiltration ceilings or repeated-jailbreak thresholds. Equally, the context layer counters malicious interaction patterns via behavioral analysis and bot detection, monitoring timing, cadence, and input regularities to distinguish genuine users from automated probing or coordinated attacks. By integrating semantic and behavioral signals, the GAF defends against adversaries who exploit not only language content but also interaction dynamics, and it tracks agent planning loops and escalating tool usage across turns.

\end{enumerate}

Table~\ref{tab:gaf-layers} summarizes the five security layers of GAF and their attack detection capabilities, showing the Generative AI (GenAI) attacks and non-GenAI attacks that can be detected by each layer.

\vspace{0.5cm}

\begin{table}[h]
\centering
\begin{tabular}{|p{0.2\textwidth}|p{0.35\textwidth}|p{0.35\textwidth}|}
\hline
\textbf{Layer Name} & \textbf{Non-GenAI Attacks} & \textbf{GenAI Attacks} \\
\hline
Network Layer & DDoS attacks, IP-based threats, TLS/SSL attacks, unauthorized access. & Prompt flooding.\\
\hline
Access Layer & Session hijacking, privilege escalation, misuse of sensitive capabilities. & AI Agent tool access manipulation. \\
\hline
Syntactic Layer & SQL injection, XSS, web application vulnerabilities. & Prompt encodings and obfuscations. \\
\hline
Semantic Layer & & Context independent GenAI attacks like DAN, DEV mode, Best-of-N jailbreak. \\
\hline
Context Layer & & Context dependent attacks like Crescendo attacks, Echo Chamber techniques, multi-turn jailbreaks. \\
\hline
\end{tabular}
\caption{Summary of GAF Security Layers and Attack Detection Capabilities}
\label{tab:gaf-layers}
\end{table}

\subsection{Extending the OSI model to LLMs}

The OSI reference model \cite{iso1994osi} is a foundational abstraction for layered communication. By decomposing systems into seven canonical layers with clearly defined functions and protocol data units (PDUs), it provides a common language for specifying protocols, reasoning about interoperability, and locating security and reliability boundaries in a deterministic, rule-based setting.

Large language models (LLMs) extend human-computer interaction beyond these assumptions. Interaction occurs through natural language, whose interpretation is probabilistic and context-dependent rather than fully prescribed by protocol syntax. This introduces a distinct interpretive boundary at the level of semantics, where failures arise from meaning manipulation (e.g., prompt injection, semantic poisoning, multi-turn exploitation). To capture this phenomenon while preserving OSI discipline, we propose a Semantic Layer (Layer 8) as a non-normative extension above the application layer; PDUs for Layers 5-8 remain "Data," consistent with X.200 usage (see Table~\ref{tab:osi_model}). Additionally, AI-to-AI exchanges (e.g., agent-to-agent or tool-calling over streams) still traverse application protocols but depend on shared semantics and intent; placing semantics above L7 reflects that failures stem from meaning rather than syntax, echoing language-based information flow control ideas \cite{sabelfeld2003language}.

The proposed L8 layer function is the interpretation and transformation of semantic content (natural-language instructions and unstructured text) into machine‑actionable representations and outputs; probabilistic inference conditioned on conversational context and knowledge; mediation between language semantics and application/service interfaces.

\begin{table}[hbt!]
\centering
\caption{The OSI Model with a proposed Semantic Layer (L8).}
\label{tab:osi_model}
\begin{tabular}{@{}ll@{}}
\toprule
\textbf{Layer} & \textbf{Protocol Data Unit (PDU)} \\
\midrule
L8. Semantic (proposed) & Data \\
L7. Application & Data \\
L6. Presentation & Data \\
L5. Session & Data \\
\midrule
L4. Transport & Segment \\
L3. Network & Packet, Datagram \\
L2. Data Link & Frame \\
L1. Physical & Bit, Symbol \\
\bottomrule
\end{tabular}
\end{table}

\subsection{Governance \& Compliance Mapping}

GAF controls align with governance frameworks that emphasize risk identification, measurement, and management, as summarized in Table~\ref{tab:gaf-governance-mapping}.

\begin{table}[h]
\centering
\begin{tabular}{@{}p{0.3\textwidth}|p{0.65\textwidth}@{}}
\toprule
\textbf{Framework area} & \textbf{GAF alignment} \\
\midrule
NIST AI RMF (Govern/Map) & Document policies, roles, and system context; map assets, threats, and impacts across GAF layers \\
\hline
NIST AI RMF (Measure) & Use red-team suites and telemetry (Table~\ref{tab:gaf-layers}) to quantify effectiveness and error rates \\
\hline
NIST AI RMF (Manage) & Enforce policies inline; operate incident response via logging, cut-offs, and escalation \\
\hline
ISO/IEC JTC 1 AI guidance & Align identity, logging, data handling, and security controls with organizational standards \\
\bottomrule
\end{tabular}
\caption{Mapping GAF to governance frameworks (illustrative).}
\label{tab:gaf-governance-mapping}
\end{table}

\subsection{What are the differences between GAF and WAF?}

The Generative Application Firewall (GAF) draws inspiration from the Web Application Firewall (WAF) but operates in distinct domains. WAFs secure structured, deterministic web interactions, while GAFs protect open-ended, language driven conversational systems. Key differences justify a separate architectural layer:

\begin{itemize}

\item \textbf{Language understanding:} WAFs do not interpret natural language. They were designed for structured, rule-based web applications, relying on pattern matching to identify threats: via strings, payload shapes, or regular expressions. These methods are not appropriate for natural language applications. A GAF must understand semantics. It needs to detect manipulative phrasing, indirect attacks, jailbreaks, or subtle attempts to extract sensitive information, even when no explicit signature is present.

\item \textbf{Context tracking:} WAFs evaluate each request in isolation, without memory of previous interactions. But generative applications are cumulative by nature. Users build up state and intent over many turns. Attacks may unfold step by step, with each individual prompt appearing harmless. A GAF must maintain session memory, reason over dialogue history, and detect escalation strategies or context-dependent violations that only emerge over time.

\item \textbf{Real-time protocols:} WAFs assume a stateless HTTP model of short, transactional, request-response cycles. But many LLM applications use streaming protocols such as WebSocket, Server-Sent Events, or gRPC. These persistent connections fall outside a WAF's scope. A GAF must operate natively within real-time interactions, staying in-session to observe and control prompts and completions as they stream in and out.

\item \textbf{Content redaction:} Traditional WAFs make binary decisions: allow or block. Generative applications require finer-grained controls. Often, only part of a model's output violates policy, for example, a leaked phone number or mention of a restricted term. Blocking the entire response would harm usability. A GAF must be able to redact selectively, mask confidential information, and cut off completions mid-generation to enforce policy in real time without degrading the user experience, and, beyond allow/block, support Redirect (soft guardrail) and Alert for fine-grained, context-aware enforcement.

\item \textbf{Interaction patterns:} WAFs are built for web interactions consisting in short, discrete actions like clicks or form submissions. LLM applications rely on open-ended conversations, where users send free-text prompts over time. This difference in cadence and structure means malicious behavior takes a different form. A GAF must model and detect threats based on how users interact in dialogue, not how they navigate a website.

\end{itemize}

\subsection{GAF Architecture}

Like a WAF, all traffic routes through GAF, intercepting incoming and outgoing flows. Plugins provide necessary protection. This integrates rather than eliminates other security layers, covering levels defined in Section~\ref{sec:layers}.

This framework proposes use cases and a high-level architecture flow. Organizations should adapt it to their needs and existing architecture. 

\begin{figure}[t]
  \centering
  \includegraphics[width=\textwidth]{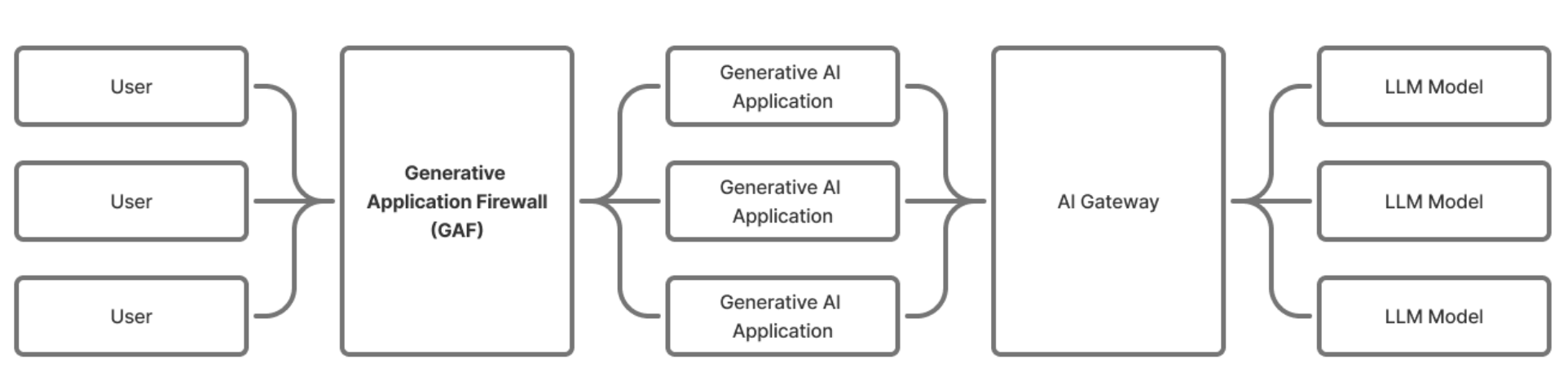}
  \caption{GAF as centralized enforcement across Network, Access, Syntactic, Semantic, and Context layers.}
  \label{fig:arch}
\end{figure}

GAF can be deployed (a) as an inline reverse proxy in front of the application, (b) as an AI-gateway sitting between the app and LLM providers/tools, or (c) as a sidecar per service in service-mesh environments. Inline modes simplify centralized policy; sidecars ease per-service autonomy.

\begin{figure}[t]
  \centering
  \includegraphics[width=\textwidth]{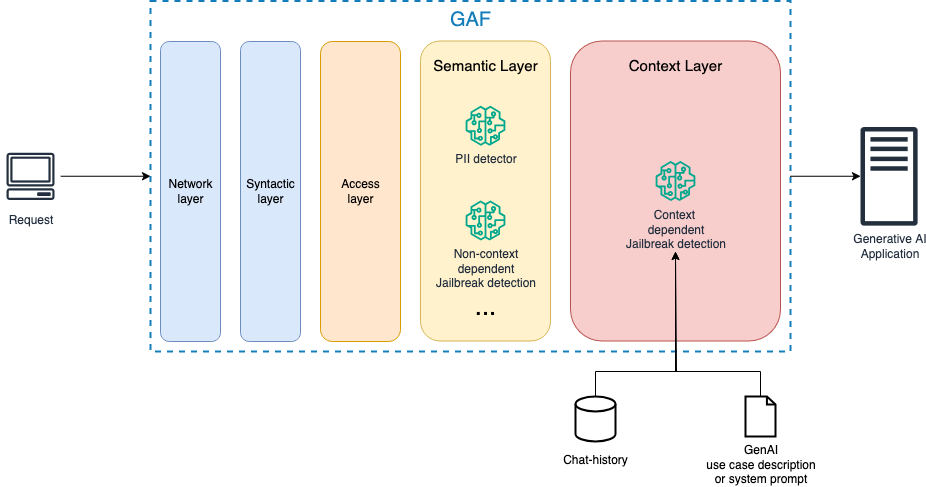}
  \caption{The context layer is the most difficult part in defining a proper GAF architecture}
  \label{fig:context-layer}
\end{figure}

The primary challenge in robust GAF architecture design resides in the Context Layer. The Semantic Layer relies on classifiers or patterns for non-context-dependent attacks, but the Context Layer addresses threats requiring deep understanding of conversational history and model use cases. Detection often requires another LLM or specialized model to analyze history and model function. Figure~\ref{fig:context-layer} illustrates this architecture, essential for multi-turn and context-dependent threats.

\subsection{Operational Performance \& Real-Time Enforcement}

GAF must balance safety enforcement with user experience. Inline controls should add minimal overhead during normal use and retain deterministic control under adversarial load. To keep GAF performant, we define a four-phase inline loop (Admission, Generation, Intervention, Post-action):
\begin{enumerate}
\item Admission: evaluate identity/session, rate limits, and basic syntactic guards before invoking the model or tools, and enforce tool-call allow/deny and scopes.
\item Generation: forward allowed requests with streaming hooks; accumulate token-level signals; monitor tool output streams.
\item Intervention: on policy triggers, selectively redact tokens or tool outputs, redirect to safe fallbacks, or terminate the stream with a well-formed close.
\item Post-action: emit structured logs/metrics and update longitudinal state for future turns, and record agent/tool-call telemetry for longitudinal analysis.
\end{enumerate}

In this loop, GAF supports granular policy actions—Block (reject a request/tool call or cut off a stream), Redact (mask spans in model or tool output), Redirect (soft guardrail to a safe response or fallback workflow), and Alert (log/notify without blocking)—to enable fine-grained, context-aware enforcement.
To measure the performance impact of GAF, we define the added latency as \(\Delta = \mathrm{p95}(t_{\mathrm{with\ GAF}}) - \mathrm{p95}(t_{\mathrm{baseline}})\), where \(t\) denotes the observed end-to-end latency for a single request. Specifically, \(t\) refers to the time taken for request setup (from user prompt submission until the LLM begins output).

During Generation and Intervention, guarantee deterministic termination and well-formed HTTP/WebSocket framing. Prefer selective token redaction over full blocking when safe, and document client-visible behaviors (e.g., cut-off markers and retry guidance) so integrators can handle them predictably.

When latency budgets are at risk or detectors degrade, fall back to conservative policies or human review, prefer fail-safe defaults for sensitive tools, and raise clear operator alerts rather than failing open.

In Post-action and across releases, emit structured logs for allow, block, redact, and terminate decisions with policy IDs, rationales, and conversation IDs. Maintain explicit SLOs for latency, error rates, and false-positive/negative rates, and validate them continuously with red-team evaluation suites and replay traffic; use these signals to tune policies and detect regressions.

\subsection{GAF Rating System}

Inspired by Tim Berners-Lee's 5-star Open Data,\footnote{\url{https://5stardata.info/en/}} this framework introduces a 5-star rating system for GAF implementations. The rating directly corresponds to the five security layers defined in Section~\ref{sec:layers}: each star represents full coverage of one layer, progressing from foundational network controls (1 star) through access, syntactic, and semantic layers, up to comprehensive context-aware protection (5 stars). This cumulative structure reflects the defense-in-depth principle---higher ratings require all lower-layer protections to be in place. Organizations can use this system to benchmark their current posture, identify gaps, and chart a maturity path toward comprehensive GenAI security.

% Bigger stars using \Large or \scalebox
\newcommand{\starrow}[1]{
  \begin{adjustbox}{max width=1.0\linewidth}
  \Large
  \foreach \x in {1,...,5} {
    \ifnum\x>#1
      \textcolor{gray!50}{\ding{72}} % empty star
    \else
      \textcolor{yellow!80!black}{\ding{72}} % filled star
    \fi
  }
  \end{adjustbox}
}

\vspace{0.5cm}

{\centering
\rowcolors{2}{gray!10}{white}
\begin{tabularx}{\textwidth}{>{\centering\arraybackslash}m{0.1\textwidth}
                >{\centering\arraybackslash}m{0.08\textwidth}
                >{\arraybackslash}X
                >{\arraybackslash}X
                >{\arraybackslash}X}
\rowcolor{gray!30}
\textbf{Stars} & \textbf{Layer} & \textbf{Protection} & \textbf{Controls} & \textbf{Attacks} \\

\starrow{1} & Network & Basic network-level filtering and access control & IP filtering, Rate and Token level Limiting, DDoS protection & DoS, IP-based threats, Prompt flooding, TLS/SSL attacks \\

\starrow{2} & Access & User and Agent authentication and authorization & Role-based access control, Session management, Policy enforcement & Session hijacking, Privilege escalation, Tool access manipulation \\

\starrow{3} & Syntactic & Syntactic analysis of inputs and outputs & Request/response filtering, Protocol validation, Format checks & SQL injection, XSS, Web vulnerabilities, Streaming protocol violations, Prompt encodings \\

\starrow{4} & Semantic & GenAI non-context dependent attacks & Prompt and context analysis, Semantic understanding, Jailbreak detection & DAN, DEV mode, Best-of-N jailbreak, Data exfiltration, PII, Prompt injection \\

\starrow{5} & Context & GenAI context dependent attacks & Multi-turn analysis, Conversation tracking, Semantic understanding & Echo-Chamber, Crescendo \\

\end{tabularx}
}

\vspace{0.5cm}

To comprehensively assess their systems, organizations should employ a reproducible test suite that covers both non-GenAI and GenAI attack families as outlined in Table~\ref{tab:gaf-layers}. This evaluation should extend beyond basic functionality to include measurements of log completeness and latency overhead under load. In particular, organizations should use a representative red-teaming corpus—incorporating both single-turn jailbreaks and multi-turn adversarial strategies—to thoroughly evaluate semantic and contextual coverage. During this process, precision/recall and false-positive rates should be reported, and p95 latency overhead introduced by GAF should be measured both at steady state and under adversarial load. Together, these practices ensure test coverage and performance are aligned with the security needs and latency budgets appropriate for the application domain.

\subsection{Maturity Path (Adoption Scenarios)}

\textbf{2-star baseline (mid-size enterprise chatbot).} Deploy GAF inline with identity integration to enforce roles and tool scopes; add rate limiting and basic syntactic validation; enable request/decision logging and start building a red-team corpus for recurring evaluation.

\textbf{Path toward 5-star (regulated institution).} Add semantic filters and selective redaction; adopt streaming cut-off for policy violations; integrate a context monitor to detect escalation across turns with thresholds and human-in-the-loop escalation; expand attack test suites and link controls to governance requirements and audit reporting.

\section{Guardrails vs. Prompt Guard vs. AI Gateway vs. GAF}\label{sec:definitions}

AI security vocabulary continues to evolve, with terms often overlapping. This section provides working definitions for common concepts in LLM-based systems:

\begin{itemize}
\item \textbf{Prompt Guard (or Prompt Shield)} \cite{inan2023llamagaurd}: A model that inspects prompts before reaching the LLM to identify and block injections and restricted topics.

\item \textbf{Guardrails} \cite{ayyamperumal2024guardrails}: Constraints that define model allowances, implemented via templates, filters, or validation.

\item \textbf{AI Gateway} \cite{neuraltrust-ai-gateway}: Middleware between application and LLM provider, handling routing, limits, logging, and caching. It decouples components for scalability, similar to API gateways in microservices.

\item \textbf{Generative Application Firewall (GAF)}: A dedicated security layer for LLM applications between user and application. It enforces policies across inputs, outputs, and interactions, understanding context and behavior. GAF orchestrates Prompt Guards, Guardrails, and components for stack-wide protection.
\end{itemize}

\section{Conclusion}

This paper introduces the Generative Application Firewall (GAF) as a new architectural layer addressing generative AI security challenges. It outlines a four-layer model—Network, Syntactic, Semantic, and Context—building on established principles while defending against semantic and conversational threats in agents and GenAI applications. It also proposes a 5-star rating system to standardize GenAI security assessment. GAF orchestrates existing tools into a unified defense, leveraging community knowledge to mitigate emerging risks.

Artificial Intelligence transforms the security landscape. While offering unprecedented utility, it introduces new risks. A dedicated layer like GAF enables comprehensive mitigation. Specific use cases may require tailoring, but pursuing a 5-star GAF rating establishes the standard for safe, responsible AI deployment.

\section*{Acknowledgements}

This work was funded by the European Union under the Horizon Europe EIC Accelerator, grant agreement No 101247606. Views and opinions expressed are however those of the author(s) only and do not necessarily reflect those of the European Union or the European Innovation Council and SMEs Executive Agency (EISMEA). Neither the European Union nor the granting authority can be held responsible for them.

\begin{center}
  \includegraphics[height=1.2cm]{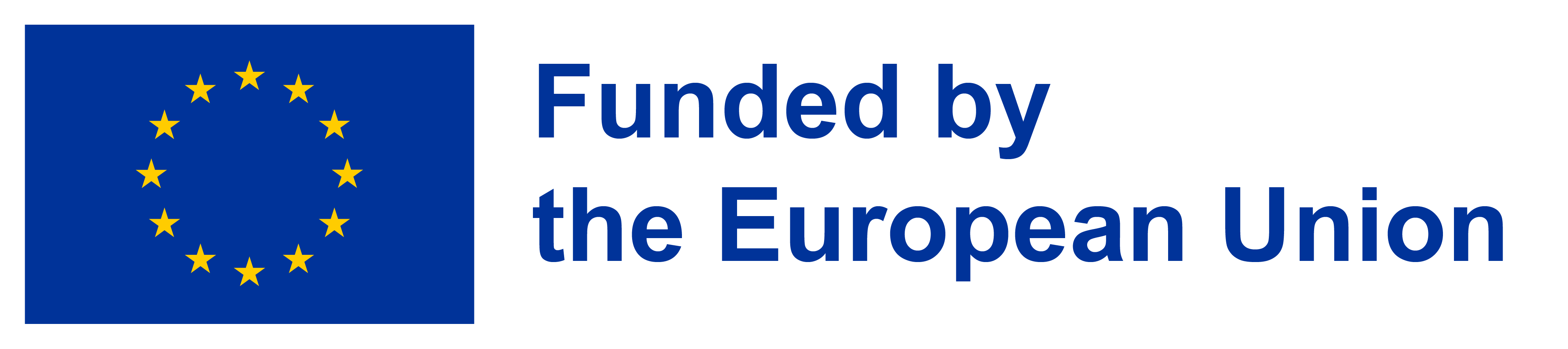}
\end{center}
\vspace{1.5mm}

\bibliographystyle{tmlr}
\bibliography{references}

\end{document}